\documentstyle[tighten,aps]{revtex}
\begin{document}
\draft
\title{Anomalous microwave response of high-temperature superconducting thin-film microstrip resonator in weak dc magnetic fields}
\author{X. S. Rao$^{*}$, ~C. K. Ong, ~B. B. Jin, ~C. Y. Tan, ~S. Y. Xu, ~P. Chen, ~J. Li, ~ and Y. P. Feng}
\address{Centre for Superconducting and Magnetic Materials
and Department of Physics, \\
National University of Singapore, Singapore 119260}
\footnotetext{$^*$ Corresponding author. Tel:+65-8742615; fax:+65-7776126; e-mail: scip7099@leonis.nus.edu.sg}
\date{\today}
\maketitle
\begin{abstract}
We have studied an anomalous microwave (mw) response of superconducting YBa$_{\rm 2}$Cu$_{\rm 3}$O$_{{\rm 7}-\delta}$ (YBCO) microstrip resonators in the presence of a weak dc magnetic field, H$_{\rm dc}$. The surface resistance (R$_{\rm s}$) and reactance (X$_{\rm s}$) show a correlated non-monotonic behaviour as a function of H$_{\rm dc}$. R$_{\rm s}$ and X$_{\rm s}$ were found to initially decrease with elevated H$_{\rm dc}$ and then increase after H$_{\rm dc}$ reaches a crossover field, H$_{\rm c}$, which is independent of the amplitude and frequency of the input mw signal within the measurements. The frequency dependence of R$_{\rm s}$ is almost linear at fixed H$_{\rm dc}$ with different magnitudes ($<$H$_{\rm c}$, $=$H$_{\rm c}$ and $>$H$_{\rm c}$). The impedance plane analysis demonstrates that r$_{\rm H}$, which is defined as the ratio of the change in R$_{\rm s}$(H$_{\rm dc}$) and that in X$_{\rm s}$(H$_{\rm dc}$), is about 0.6 at H$_{\rm dc}$$<$H$_{\rm c}$ and 0.1 at H$_{\rm dc}$$>$H$_{\rm c}$. The H$_{\rm dc}$ dependence of the surface impedance is qualitatively independent of the orientation of H$_{dc}$. 
\end{abstract}
\pacs{PACS number(s): 74.25.Nf, 74.76.Bz, 74.60.Ge, 74.60.Ec.}

\section{Introduction}

In general, it is expected that the surface impedance of high-temperature superconducting (HTS) thin films will increase as a function of applied dc or mw field. This nonlinearity was believed to be mainly contributed by weak links and vortex dynamics in the HTS films \cite{Hylton88,Portis89,Halbritter90,Coffey91,Nguyen93,Shridhar94,Nguyen95,Willemsen95,Golosovsky95,Golosovsky96,Cohen97,Tsindlekht97}. Besides, other sources of nonlinearity, such as intrinsic pairbreaking, local heating of grain boundaries and thermal switch of superconducting grains, were also introduced to explain the diverged experimental data\cite{Lam92,Wosik97,Hein97a}.  

Recently, several anomalous microwave responses, i.e. surface resistance R$_{\rm s}$ and (or) surface reactance X$_{\rm s}$ change non-monotonically with applied mw or dc magnetic field, were observed \cite{Choudhury97,Hein97,Kharel98,Kharel99}. Choudhury $et$ $al$. reported an anomalous field dependence of surface resistance R$_{\rm s}$ of a YBCO suspended line resonator in the presence of a weak, perpendicular dc magnetic field at 10K. They found a sharp decrease of R$_{\rm s}$ at small H$_{\rm dc}$ ($\sim$5 Gauss) with subsequent increase at higher fields. Hein $et$ $al$. observed a correlated reduction in R$_{\rm s}$ and X$_{\rm s}$ of epitaxial YBCO films in both dc and mw magnetic fields of the scale 20 mT at frequencies f=8.5 and 77 GHz and temperatures T=77 and 4.2K, respectively. The field effects on R$_{\rm s}$ and X$_{\rm s}$ were found to correlate within the framework of the two-fluid model (TFM). The reduction in R$_{\rm s}$ and X$_{\rm s}$ was attributed to magnetic field-induced suppression of the spin-flip scattering rate. Similar effects were also reported by Kharel and his colleagues. Besides correlated changes in R$_{\rm s}$ and X$_{\rm s}$, uncorrelated non-monotonic field dependence was also observed in their measurement of YBCO coplanar resonators at 8.0GHz and different temperatures (15, 35 and 75K). 
As far as we know, the origins of these anomalous microwave responses are still pending and the experimental data are very limited. More experimental investigations are needed to give enough clues for learning the mechanisms underneath the anomalous effects.

In this work, we employed a microstrip resonator technique to measure R$_{\rm s}$ and X$_{\rm s}$ of YBCO thin films in dc magnetic fields H$_{\rm dc}$ up to 200 Gauss at 77K. It was observed that R$_{\rm s}$ decreases as H$_{\rm dc}$ raises from zero and this is followed by a monotonically increase of R$_{\rm s}$ at higher H$_{\rm dc}$. Correlatively, X$_{\rm s}$ show a similar behaivior. The effects were investigated at different mw amplitudes and frequencies in different dc field alignments. The experimental data have been examined with several theoretical models.

\section{Sample preparation}

The double-sided YBCO thin films were prepared by on-axis pulsed-laser deposition (PLD) technique on polished LaAlO$_{\rm{3}}$ (100) substrates with the size of 15$\times$10$\times$0.5 mm$^{\rm 3}$. The films are 400nm thick and show a strong preferential orientation with the c-axis perpendicular to the film surface. The AFM results show that the grain sizes of both faces of the double-sided films are around 1$\mu$m$\times$1$\mu$m and the root-mean-square (rms) surface roughness of both faces in a 50$\mu$m$\times$50$\mu$m area are similar, around 170 nm while their mean roughness are also similar, which are around 130 nm. The dc critical current density J$_{\rm c}$ is in excess of 10$^{\rm 6}$A/cm$^{\rm 2}$ at 77K and the transition temperature T$_{\rm c}$ is around 90K with a narrow transition width, 0.5$\sim$1K. Employing a dielectric resonant cavity \cite{Ong99}, the surface resistances of the films were measured at 10.66GHz, 77K before patterning.  The resultant R$_{\rm s}$ are around 1 m$\Omega$ and the difference of the surface resistance between the two sides is smaller than 5\%. 

The design of the microstrip resonator was done with the help of a commercial full wave electromagnetic simulator, IE3D, and the quasi-TEM analytical formulations \cite{Gupta79} for microstrip resonator. The resonator geometry is chosen so that it has a characteristic impedance of 50$\Omega$. The meandering microstrip has a width of 169 $\mu$m and is coupled with the outer circuits by a capacitive gap of 500 $\mu$m. The traditional wet etching method was employed to fabricate the microstrip resonator. We have checked the geometry of the resonator after the etching process and found the actual strip width is around 4$\mu$m narrower than the designed value. 

\section{Experimental results}

The microwave responses of the resonators were measured at 77K using a vector network analyzer and the dc magnetic field was applied by a copper solenoid. There is no remanent field in the measurements since all the experimental set-ups are made of Teflon except the copper solenoid.  The changes in surface impedance associated with the application of dc magnetic field are extracted from the measured quantities as follow \cite{Willemsen97}:
\begin{equation}
\Delta {\rm Z}_{\rm s}({\rm H}_{\rm dc})={\rm Z}_{\rm s}({\rm H}_{\rm dc})-{\rm Z}_{\rm s}({\rm 0})=\Gamma [ \Delta {\rm f}_{\rm -3dB}({\rm H}_{\rm dc})-\Delta {\rm f}_{\rm -3dB}({\rm 0})+ i2({\rm f}_{\rm 0}({\rm 0})-{\rm f}_{\rm 0}({\rm H}_{\rm dc}))]
\end{equation}
where $\Gamma$ is a geometric factor determined from the sample dimensions, f$_{\rm 0}$ and $\Delta$f$_{\rm -3dB}$ are the resonance frequency and the $-$3dB (half power) bandwidth of the resonance peak, respectively. The loaded quality factor Q$_{\rm L}$ of the resonator can be easily obtained from f$_{\rm 0}$ and $\Delta$f$_{\rm -3dB}$, Q$_{\rm L}$=f$_{\rm 0}$/$\Delta$f$_{\rm -3dB}$. We replace Q$_{\rm L}$ with unloaded quality factor because the coupling is very weak (insertion loss$>$30dB). 

Figure 1(a) shows the typical data for $\Delta$f$_{\rm -3dB}$, which is proportional to R$_{\rm s}$, of the fundamental resonant peak as a function of applied dc field H$_{\rm dc}$. The input mw power was fixed at $-$10dBm and the sample was initially cooled in a zero-field state where the earth magnetic field ($\sim$0.3 Gauss) was neglected. We do not give absolute values of R$_{\rm s}$ and X$_{\rm s}$ because $\Gamma$ is difficult to be determined for the current in the microstrip is highly non-uniform \cite{Sheen91,Dahm97}. Fortunately, only the functional dependence of $\Delta$Z$_{\rm s}$ and the ratio of $\Delta$R$_{\rm s}$ and $\Delta$X$_{\rm s}$ are of importance to learn the mechanisms of microwave nonlinearity and they can be determined without the geometric factor. For the case H$_{\rm dc}$ was applied parallel to the $c$-axis of the film (namely, perpendicular to the surface of the film), R$_{\rm s}$ drops dramatically as H$_{\rm dc}$ is increased from zero. After passing through a minimum, R$_{\rm s}$ begin to increase monotonically when H$_{\rm dc}$ is further raised. The minimum depth is 23\% of the R$_{\rm s}$ value in zero dc field and the crossover field (H$_{\rm c}^{\|}$) is 4.5 Gauss, which is of the same order of that reported by Choudhury $et$ $al$. \cite{Choudhury97}. The shift of the resonant frequency, f$_{\rm 0}$, shown in Figure 1(b) reveals that the change of X$_{\rm s}$ is correlated with that of R$_{\rm s}$. The resonant peak initially shifts to higher frequency (X$_{\rm s}$ decreases, correspondingly) and then shifts down. The H$_{\rm c}^{\|}$ value determined from X$_{\rm s}$(H$_{\rm dc}$) is the same as that from R$_{\rm s}$(H$_{\rm dc}$) within 1.5 Gauss.  

The general features observed above were presented again for H$_{\rm dc}$$\bot$$c$ while the crossover field H$_{\rm dc}^{\bot}$ is of the order 80$\sim$90 Gauss. The difference between H$_{\rm dc}^{\bot}$ and H$_{\rm dc}^{\|}$ may come from the serious demagnetizing effect and the anisotropy of the HTS film. The insets in Fig.1 show the curves normalised with H$_{\rm dc}^{\bot}$ and H$_{\rm dc}^{\|}$, respectively. The two curves almost collapse to one. The implication of this result will be discussed below. 

Similar effects are observed as the input power is varied. Fig.2 shows the $\Delta$f$_{\rm -3dB}$ versus H$_{\rm dc}$ with different orientations when the input mw power is fixed at $-$20dBm. No shift of the crossover field H$_{\rm c}$, which represents H$_{\rm dc}$$\bot$$c$ or H$_{\rm dc}$$\|$$c$, at different mw power levels is observed within the accuracy of the measurement. It means that H$_{\rm c}$ is independent of the amplitude of the input microwave signal. However, the decrease in R$_{\rm s}$ is more prominent than that of P=$-$10dBm. The minimum depth in this case is as large as 43\% of the zero-dc field R$_{\rm s}$ value. 

The same measurements were also carried out for different resonant modes (second and third harmonics), shown in Fig.3(a). The field dependencies of R$_{\rm s}$ for different resonant modes are qualitatively the same. R$_{\rm s}$ first decreases with H$_{\rm dc}$ and then increase after H$_{\rm dc}$ reaches the crossover field H$_{\rm c}$. The crossover fields H$_{\rm c}$ measured from the second and third modes are almost the same as that from the fundamental mode. Thus H$_{\rm c}$ is also independent of the frequency of the input mw signals. As is well known, the frequency dependence of R$_{\rm s}$ is an important key in the determination of the microwave loss mechanism. To shed some light on the microwave loss mechanism in the applied dc magnetic field with different magnitudes, we chose three typical H$_{\rm dc}$: zero dc field, H$_{\rm c}$ and a field larger than H$_{\rm c}$ to plot the $-$3dB bandwidth as a function of the harmonic numbers, as shown in Fig.3(b). Nearly linear frequency dependencies of R$_{\rm s}$ are found for all the three values of H$_{\rm dc}$ and even the slopes of the three curves are almost independent of H$_{\rm dc}$. The similarity in the frequency dependence of R$_{\rm s}$ would suggest the microwave loss mechanisms at the three dc fields are essentially the same as each others.  The linear frequency dependence rules out the loss mechanisms with f$^{\rm 2}$ dependence, such as two fluid model \cite{Lancaster97}, BCS theory \cite{Turneaure91} and weakly coupled grain model \cite{Hylton88,Nguyen93}. And it implies that the hysteretic losses due to pinning and nucleation of Josephson fluxons \cite{Portis89} may play an important role in the loss mechanism. 

Although the loss mechanism is almost same in our measurements, the effect of H$_{\rm dc}$ on the loss mechanism is totally different for the case H$_{\rm dc}$$<$ H$_{\rm c}$ and H$_{\rm dc}$$>$H$_{\rm c}$. To illustrate the effect of the applied dc field, we adopt an impedance plane analysis, which has been proven to be a powerful approach in distinguishing between various nonlinear mechanisms of superconductors, in terms of r-parameter \cite{Golosovsky95,Halbritter97}. Here the dc field-induced changes in the surface impedance are characterised by a dimensionless r$_{\rm H}$ which is defined as the ratio of the change of surface resistance and reactance with varing H$_{\rm dc}$ at fixed input mw powers, i.e., r$_{\rm H}$ = $\Delta$R$_{\rm s}$(H$_{\rm dc}$)/$\Delta$X$_{\rm s}$(H$_{\rm dc}$). Figure 4 demonstrates $\Delta$f$_{\rm -3dB}$ versus f$_{\rm 0}$ dependencies at P$_{\rm mw}$=$-$10dBm for varied H$_{\rm dc}$ values and for different orientations of the applied dc magnetic field. The r$_{\rm H}$-parameter can be easily extracted from the slopes of the curves (r$_{\rm H}$ is one-half of the slope). The obtained value for r$_{\rm H}$ is about 0.6 if H$_{\rm dc}$ is below H$_{\rm c}$, while r$_{\rm H}$$\sim$0.1 if the dc field is above H$_{\rm c}$. The transition of r$_{\rm H}$ happens within a narrow field range at the crossover dc field. The prominent difference in the values of r$_{\rm H}$ implies that the dc field below H$_{\rm c}$ plays a different role with that above H$_{\rm dc}$ in determining the surface impedance. It is noted that the r$_{\rm H}$$\sim$0.1 at H$_{\rm dc}$$>$ H$_{\rm c}$ is consistent with the r$_{\rm H}$ values (0.1 - 0.2) reported by other groups \cite{Cohen97,Tsindlekht97,Andreone97}. The impedance plane analysis at P$_{\rm mw}$=$-$20dBm presents a similar behavior. It shound be stressed that Fig.4 is qualitatively different from the similar analysis given in Fig.5 of Ref. 17. This suggests that the anomalous response reported by Hein $et$ $al$. and what observed in this work may come from different origins.

In addition, Fig.4 shows that for different orientations of H$_{\rm dc}$ ( H$_{\rm dc}$$\|$$c$ and H$_{dc}$$\bot$$c$ ) the two sets of data points essentially coincide with each other on the same curve. As shown in the insets of Figure 1, the R$_{\rm s}$(H$_{\rm dc}$) and X$_{\rm s}$(H$_{\rm dc}$) curves with different H$_{\rm dc}$ orientations can also be normalised to coincide with each other. Combining these observations together, we conclude that the effects of H$_{\rm dc}$ on the surface impedance are essentially independent on the orientations of the applied dc magnetic field.  

Three samples were measured in this work. Though quantitative differences are found, the features of the data are essentially the same as mentioned above. The quantitative diferences may come from the differences on growth, deposition and patterning of the films. 

\section{Discussions}

The results show that H$_{\rm c}$ is independent of both the amplitude and frequency of the applied microwave signal. It implies that H$_{\rm c}$ may be related to some characteristic parameters of the YBCO films. Since the effects of dc magnetic field change dramatically at H$_{\rm c}$, one could naturally expect that the observed H$_{\rm c}$ in the measurements is simply a manifestation of H$_{\rm c1}$, the lower critical field of the superconducting YBCO films. Below H$_{\rm c}$, the external dc field causes a decrease in R$_{\rm s}$ for some reasons. Above H$_{\rm c}$, the dc fluxons begin to penetrate into the samples and produce additional loss with a large increase of the surface impedance. This may give an qualitative description on some features of the anomalous effect mentioned above. However, we can not definitely clarify the relationship of H$_{\rm c}$ to H$_{\rm c1}$ while the origin of the drop in R$_{\rm s}$ is still not known. The conception, which says the observed H$_{\rm c}$ is a manifestation of H$_{\rm c1}$, needs to be further studied.

Recently, several groups proposed that magnetic field-induced recovery of superconductivity might account for the anomalous microwave response \cite{Hein97,Kharel98,Kharel99}. Magnetic impurities are likely to be present in most HTS materials. The interaction between localised magnetic moments of magnetic impurities and cooper pairs which are in the singlet state destroys the pair correlation and is accompanied by spin-flip scattering \cite{Kresin96}. An external magnetic field forces the localised magnetic moments to align, frustrates the spin-flip scattering, and leads to a reduction of pair breaking. According to two-fluid model the increase of pair electrons can lead to the decrease of R$_{\rm s}$ and X$_{\rm s}$. This mechanism has been proven to be effective in explaining some of the experimental results \cite{Hein97}. However, it is obvious that the feature of Fig.4 is qualitatively different from that of the similar analysis given in Fig.5 of Ref.[17]. So it is doubtful that the anomalous response we observed comes from the same origin as that in Ref.[17]. In addition, a phenomenological description is proposed in terms of this mechanism \cite{Jin99}. Simulation shows that quantitative fitting of the experimental data with this model requires a large increase of pair density ($>$50\%) with a relatively small change of external field (80$\sim$90 Gauss). So the applicability of this mechanism to the present experimental results is questionable. 

A phenomenological model of the nonlinear microwave response of a superconducting weak link was proposed by Velichko to describe the effect of both dc and mw magnetic fields \cite{Velichko99}. The results show that the value of Z$_{\rm s}$(H) of both ``the non-shunted'' and ``the shunted WL'' can fall with increased H under certain conditions. However, the observed correlated decrease in R$_{\rm s}$ and X$_{\rm s}$ still can not be explained by the model. For ``the non-shunted WL'', R$_{\rm s}$ and X$_{\rm s}$ initially increase as H raises from zero and then decrease with elevated magnetic field, which is inconsistent with our experimental observation. For ``the shunted WL'', a decrease in R$_{\rm s}$ is accompanied by an increase of X$_{\rm s}$. This is also inconsistent with our data. As shown by Herd $et$ $al$., however, if a network of weak links with a distribution of the I$_{\rm c}$R$_{\rm n}$ products was considered, the decrease of X$_{\rm s}$ with H can be expected \cite{Herd97}. It is not sure whether a network of ``the shunt WL'' can account for the decrease of both R$_{\rm s}$ and X$_{\rm s}$.

Similar effects were once observed by Thompson $et$ $al$ \cite{Thompson79}. in the measurements of ac hysteretic losses in Nb$_{\rm 3}$Ge material. With fixed ac amplitude, the ac loss was observed initially decrease and then increase with elevated dc-bias magnetic field after passing through a minimum. The decrease of ac losses was explained in terms of Abrikosov vortex-antivortex annihilation within the frame of critical state model \cite{Clem79}. One of the central features of this theory lies in an ac amplitude-dependent H$_{\rm c}$ at which ac loss reaches its minimum. While this feature was observed in Thompson's experiment, it is opposite to our observation. As we mentioned above, H$_{\rm c}$ is almost independent of the amplitude of the input mw signal. 

So far the observed anomalous microwave response can not be explained consistently by any of the proposed models. It is generally expected that both of the intrinsic and extrinsic effects in the HTS thin films may have their contributions to the microwave loss. To get a clear understanding of the anomalous effect, all these effects should be taken into account. The high value of R$_{\rm s}$ ($\sim$ 1m$\Omega$ at 10.66 GHz, 77K) shows that the films used in this work are highly granular and most likely consist of a microbridge-type weak links. The weak links may play an important role in the anomalous effects though the mechanism is not known at present. As we know, granularity has been ruled out as a general source of the anomaly observed in their experiments since their films showed well-established qualities and excellent power handling capabilities. This can also explain why Fig.4 is qulitatively different from the Fig.5 in Ref.17. This suggest that the anomalous effects may have more than one source.  

\section{Conclusions}

We have observed an anomalous microwave response in HTS thin-film microstrip resonators. 
The main results of this work are summarized as follows: (a) The surface resistance and reactance show a correlated non-monotonic behavior in the presence of a weak dc magnetic field. The R$_{\rm s}$ drops as H$_{\rm dc}$ raises from zero, after passing through a minimum, it increases gradually as H$_{\rm dc}$ is further increased. The X$_{\rm s}$ presents a similar behavior; (b) The crossover field H$_{\rm c}$ , at which R$_{\rm s}$ reaches its minimum, is independent of the amplitude and frequency of the input microwave signal within the ranges of measurements; (c) The qualitative effects of H$_{\rm dc}$ on the surface impedance are essentially independent of dc field orientation, microwave amplitude and frequency, while prominently different at H$_{\rm dc}$$<$H$_{\rm c}$ (r$_{\rm H}$$\sim$0.6) and at H$_{\rm dc}$$>$H$_{\rm c}$ (r$_{\rm H}$$\sim$0.1); (d) At a fixed H$_{\rm dc}$ and P$_{\rm mw}$, the frequency dependence of the surface resistance is almost linear. This linear frequency dependence and even the slope of R$_{\rm s}$-f are unchanged at varied H$_{\rm dc}$ magnitudes ($<$H$_{\rm c}$, $=$H$_{\rm c}$ and $>$H$_{\rm c}$). Several mechanisms were examined and they can not give a satisfactory explanation on the results. The origin of the observed anomaly is still not clear. Further efforts should be addressed to understand it fully.

\begin{figure}
\caption{ dc magnetic field dependence of (a) the $-$3dB bandwidth $\Delta$f$_{\rm -3dB}$ and (b) the resonant frequency  f$_{\rm 0}$ of the microstrip resonator working at fundamental mode with input power P$_{\rm mw}$=$-$10dBm. Circles correspond to H$_{\rm dc}$$\|$$c$ and triangles correspond to H$_{\rm dc}$$\bot$$c$. The insets show  the curves normalised with H$_{\rm dc}^{\|}$ and H$_{\rm dc}^{\bot}$, respectively.}
\label{Fig.1}
\end{figure} 
\begin{figure}
\caption{ dc magnetic field dependence of $\Delta$f$_{\rm -3dB}$ of the microstrip resonator working at fundamental mode with input power P$_{\rm mw}$=$-$20dBm. Circles correspond to H$_{\rm dc}$$\|$$c$ and triangles correspond to H$_{\rm dc}$$\bot$$c$.}
\label{Fig.2}
\end{figure}
\begin{figure}
\caption{ (a) dc magnetic field ($\|$$c$) dependence of $\Delta$f$_{\rm -3dB}$ of the microstrip resonator working at fundamental mode (circles), second (triangles) and third harmonics (squares) with input power P$_{\rm mw}$=$-$10dBm. (b) Frequency dependence of  $\Delta$f$_{\rm -3dB}$ at different dc magnetic fields.}
\label{Fig.3}
\end{figure}
\begin{figure}
\caption{ $\Delta$f$_{\rm -3dB}$ vs f$_{\rm 0}$ on varying dc magnetic field for the microstrip resonator working at fundamental mode with input power P$_{\rm mw}$=$-$10dBm. Circles correspond to H$_{\rm dc}$$\|$$c$ and triangles correspond to H$_{\rm dc}$$\bot$$c$.}
\label{Fig.4}
\end{figure}

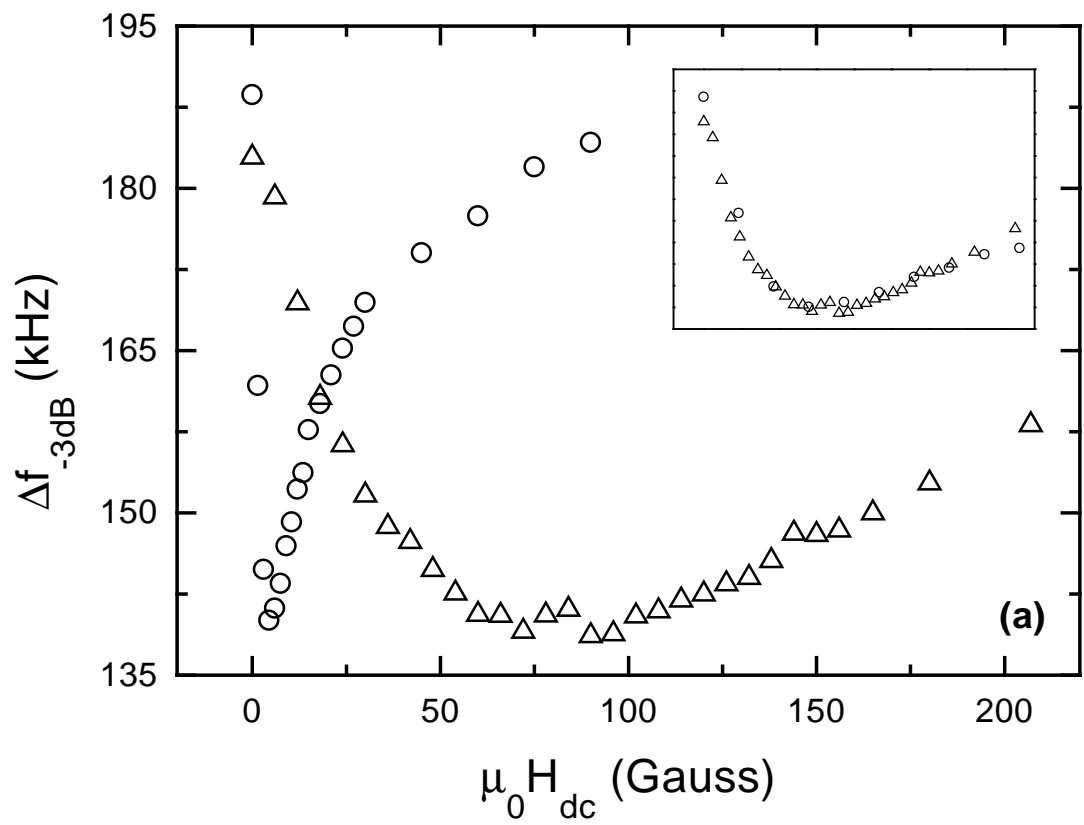
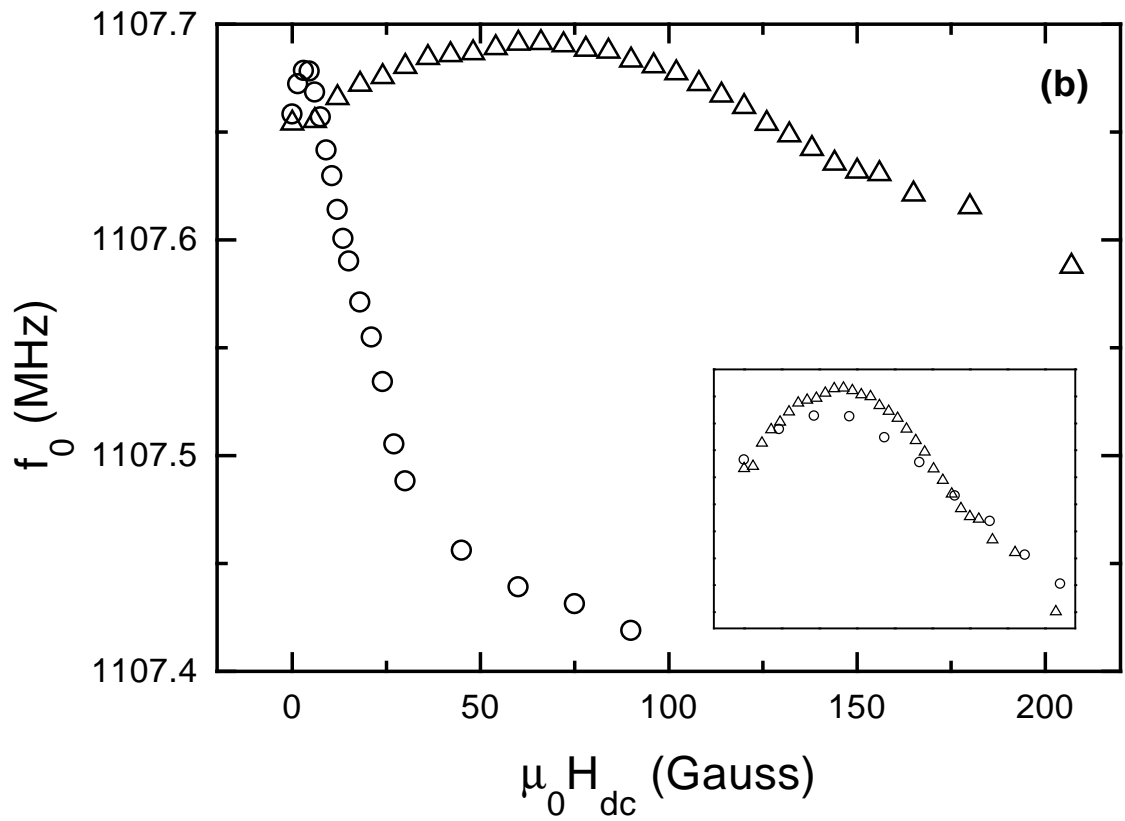
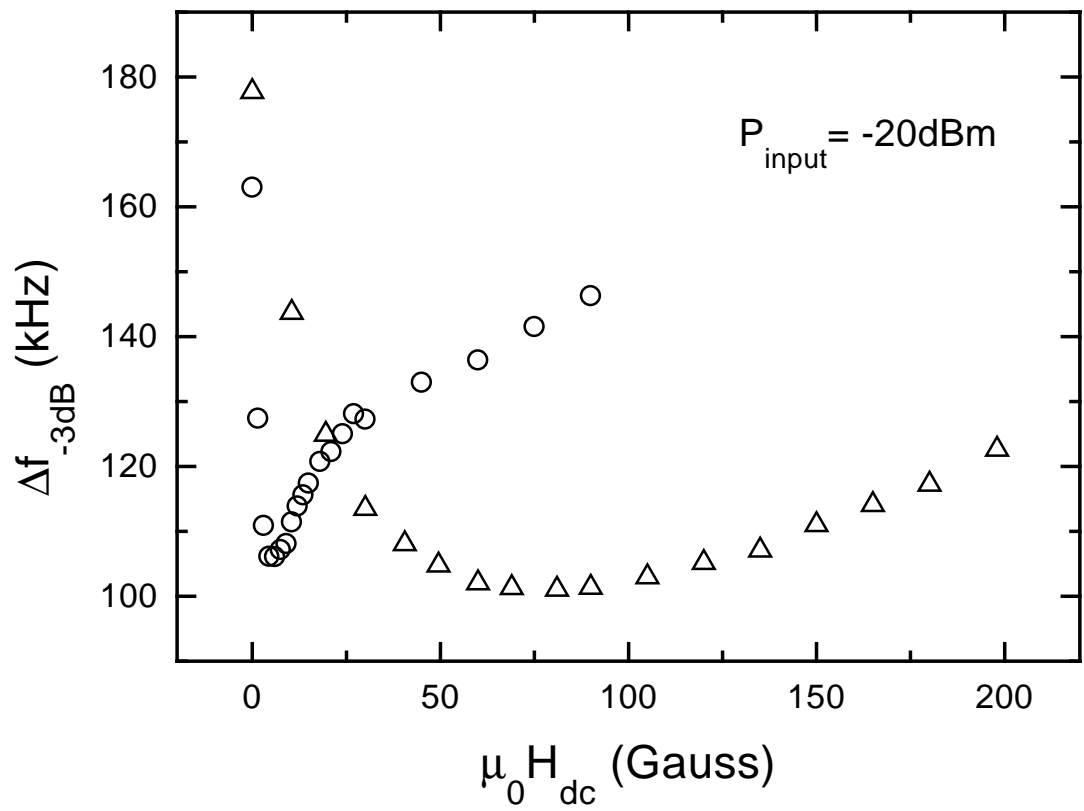
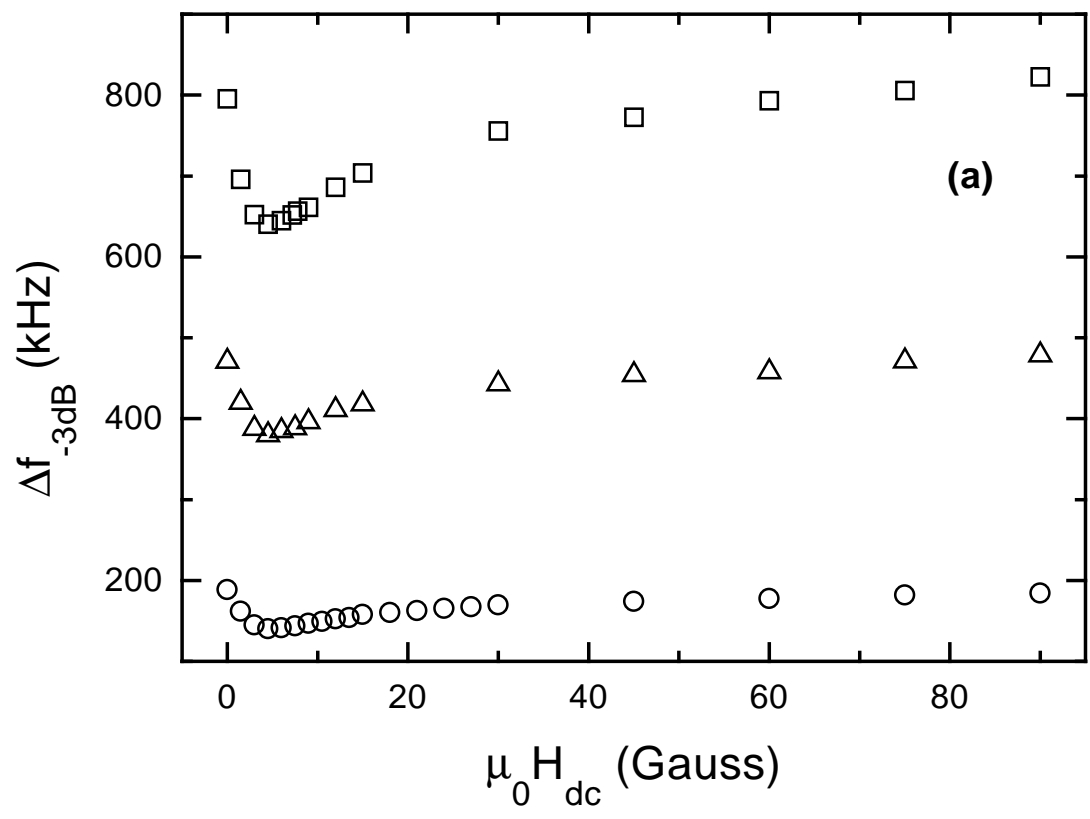
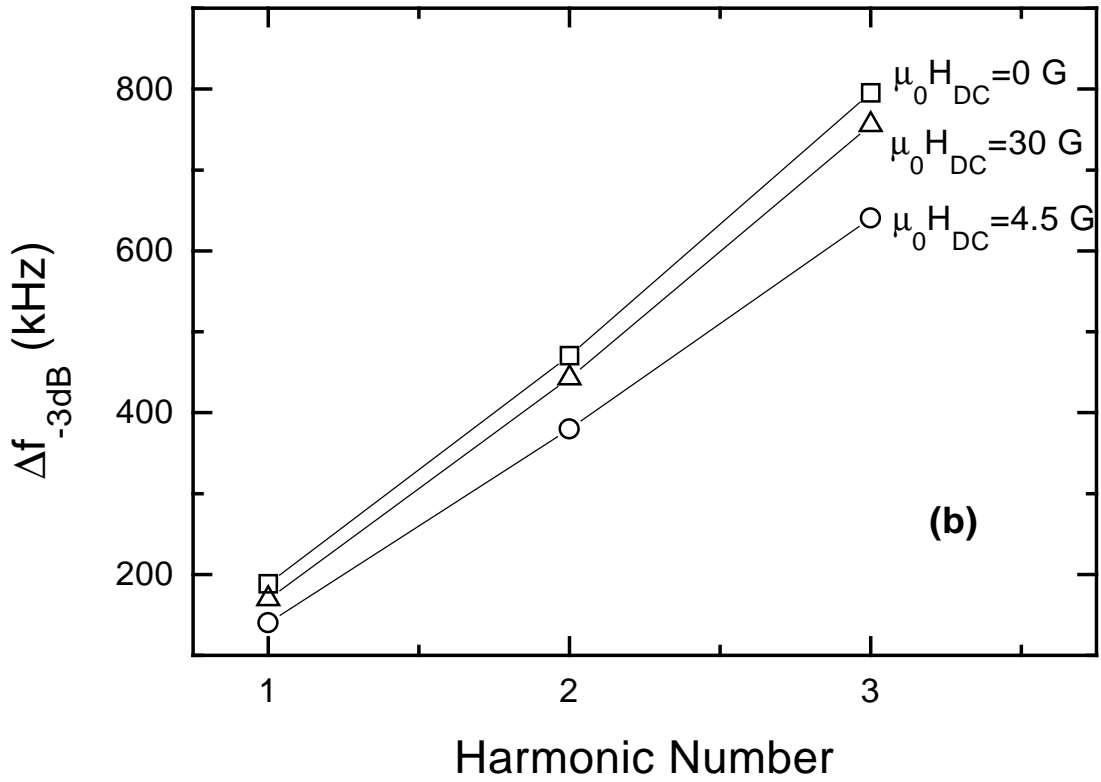
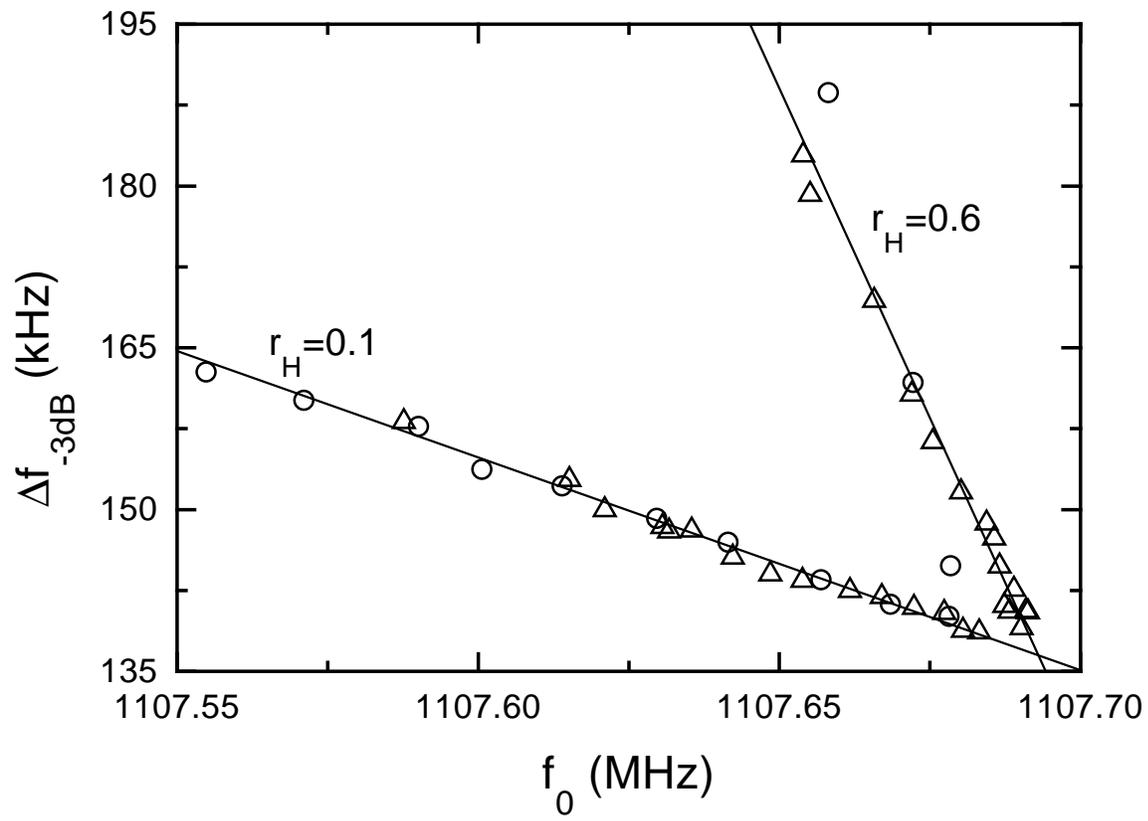
\end{document}